\documentclass[%
 reprint,
 amsmath,amssymb,
 aps,
]{revtex4-2}
\usepackage{graphicx} 
\usepackage{dcolumn}
\usepackage{bm}
\usepackage{hyperref}
\usepackage{xcolor}    

\usepackage[mathlines]{lineno}

\hypersetup{
    colorlinks=true,
    citecolor=blue,
    linkcolor=blue,
    urlcolor=blue
}

\let\oldcite\cite
\renewcommand{\cite}[1]{\textcolor{blue}{\oldcite{#1}}}

\begin{document}
\preprint{APS/123-QED}

\title{The prospect of confining the equation of state of neutron star with future mass and radius measurement}

\author{Asim Kumar Saha}%
 \email{asim21@iiserb.ac.in}
\affiliation{%
 Department of Physics,\\ Indian Institute of Science Education and Research Bhopal, Bhopal, India.}
\author{Ritam Mallick}
\email{mallick@iiserb.ac.in}
 \affiliation{Department of Physics,\\ Indian Institute of Science Education and Research Bhopal, Bhopal, India.}


\date{\today}
\begin{abstract}
Simultaneous measurement of mass and radius with high precision is essential to unravel the equation of state of matter at the centre of neutron stars. Measurement of massive pulsars indicates that the equation of state has to be stiff at low densities. The radius measurement of PSR J0030+0451 has rejected several relatively softer equations of state. In this work, an ensemble of agnostically constructed EoS was studied for the mass and radius measurement. The range of radius of neutron stars obtained from the ensemble was confined within a radius bound from 10.5 - 14.5 km. It is seen that higher masses prefer stiffer EoS. However, the slope of the speed of sound (the stiffness of the EoS) is very sensitive to the radius measurement. Assuming the radius measurement to be precise up to 2 km, then a higher radius indicates sharp stiffening of the equation of state at low density; however, it also indicates a sharp fall after the peak, indicating a relatively softer core for massive neutron stars. On the other hand, for the same precision, a lower radius measurement indicates a relatively softer equation of state.
\end{abstract}

\maketitle
\section{\label{sec:level1}Introduction}

The properties of matter (of a state) responsible for its characteristic behaviour depend on its thermodynamic variables. Experiments and theoretical understanding go hand in hand to understand the matter's properties. However, properties of matter at high temperatures and density are not easy to access as they are not found naturally on Earth. To unravel the matter properties at high temperatures, earth-based collider experiments are being performed along with improved theoretical and computational methods \cite{Huth2022, Adamczewski-Musch2019,Jiang:2022tps,PARKKILA2022137485}. The matter properties at high density are equally interesting; however, analyzing them has proved difficult till now. Earth-based experiments are still awaited, and theoretical u ApJLnderstanding at high densities is hindered by some complex problems \cite{Durante_2019, Geraksiev_2018, 2017JInst..12C6012K,Peter_Senger}. Astrophysical observation of pulsars has proved immensely important in studying matter properties at high densities \cite{LATTIMER2007109}.

Neutron stars are pulsars which emit radiation in pulses mainly in the radio waves and sometimes in x-rays or even in gamma rays \cite{Abdo_2013, Mori_2014, HEWISH1968, Ozel_review}. Pulsars can be as massive as a few solar masses (1-3 solar mass) but of a radius of 10-15 km \cite{Lattimer_Prakash_review, Lattimer_book, Glendenning:1997wn}. This makes them very dense, and the core density could be 6-8 times nuclear saturation density \cite{Lattimer_Prakash_review, Ozel_review}. This density range coincides with the high density, which cannot be directly probed experimentally, nor does it have a good theoretical description. Also, in this density range, the phase transition from confined matter to deconfined matter has been proposed, which makes them even more interesting \cite{GLENDENNING2001393,Annala2020,Staff_2006, Weber_2014}.

At lower densities, i.e. at densities below saturation density ($n \lesssim 2n_0$), effective field theory (EFTs) models have proven to be adequate to describe matter \cite{Tews_2013,Bogner_2011,Drischler_2019}. The matter at these densities can be probed in nuclear experiments allowing for a high confidence in its properties \cite{Tsang_2012, Brenden_2021, Huth2022}. EFT models break down at higher densities, so calculations based on EFTs are no longer applicable as uncertainty increases with density \cite{Tews_2018, Londardoni_2020, Leonhardt_2020}. On the other hand, at asymptotically high densities, i.e. densities much higher than those at the core of neutron stars ($n\gg n_0$), 
quarks and gluons are in a deconfined state, which enables the applicability of perturbative QCD calculations \cite{Lattimer_2004, Freedman_McLerran, Vuorinen_2003}.
At intermediate densities, i.e. at densities a few times the saturation density ($n_0$), where the cores of neutron stars are believed to exist, the properties of matter are still uncertain. The way of unravelling the matter properties at these densities can be through theoretically modelling the matter (using symmetries) and solving for NS properties. The outcome of the computational models can then be compared with observation from pulsars. The properties of matter are usually expressed as their equation of state (EoS), a relation between thermodynamic variables, which gives a mathematical description of matter with its thermodynamic properties. 
The EoS in this regime can either be built theoretically that is model specific \cite{Oertel_2017, Bastian_2021, Ishii2019} or by sampling them agnostically \cite{Greif_2019, Annala2020, Somasundaram_2023}.

NSs are one of the densest objects in the universe; however, deciphering the observational data of NS with the theories is complex. This is because most observations directly pertain to the surface of neutron stars, with the core hidden deep inside, where our interest lies \citep{Nath_2023}. 
In order to test the EoSs from either theoretical or agnostic modelling, we must model NS adhering to TOV equation \citep{Tolman_1939, OV_equation}. The physical properties of mass and radius resulting from solving TOV equations are then compared with the observations.  
The observations of NSs has been conducted by various methods i.e. radio \citep{Crawford_2006, Boyles_2013, Cromartie2020, Fonseca_2021}, electromagnetic (EM) \citep{Abbott_2017}, x-ray \citep{Miller_2019,Miller_2021, Riley_2019, Riley_2021} and gravitational wave (GW) observations \citep{Abbott_GW1, Abbott_2018}. Detection of GWs \citep{Abbott_2016, Castelvecchi2016} has opened up the possibility to constrain EoS with an added physical parameter - tidal deformability. Tidal deformability, directly related to the neutron star structure, depends heavily on the EoS being considered \citep{Hinderer_2010, Flanagan_2008, Bernuzzi2020, Lackey_2015}.

 In the last 15-20 years, the astrophysical observation of neutron stars has developed leaps and bounds. Presently, we are able to measure (although indirectly) the mass, its magnetic field, radius, rotational velocity and tidal deformability with reasonable accuracy. Pulsar mass measurements can be obtained through precise radio timings, sometimes supplemented by optical observations \citep{antoniadis2016millisecondpulsarmassdistribution}. Observations of radio timings yield measurements of orbital period and semi-major axis, which allows us to infer the unknown stellar masses of the pulsar and its companion, using post-Keplarian parameters induced by relativistic effects \citep{2004hpabookLorimer}. Phase-resolved spectroscopy can yield the binary's mass ratio, allowing for accurate mass measurements\citep{Antoniadis_2013, Tremblay_2013}. Information about the magnetic field can be estimated by considering the spin-down of NS is due to magnetic dipole radiation loss. Hence, precise measurement of period and period derivative provides information on the star's magnetic field strength \citep{Makishima}. Apart from the magnetic field, it can also provide us with the characteristic age of the pulsar\citep{Igoshev_2022}. All these measurements have proved quite decisive in effectively constraining the matter properties. 
 
Along with the measurement, agnostically constructed EoS randomizing the speed of sound has given some bound on the EoS at the intermediate densities.
The adiabatic sound speed $c^2_s = \left(\frac{\partial p}{\partial e}\right)_s$, here $s$, $p$, $e$ refers to specific entropy, pressure and energy density, respectively. Thermodynamic stability and causality bounds the sound speed as $0 \leq c^2_s \leq 1$ \citep{Chimowitz, HARTLE1978201}. It determines the maximum mass that an EoS can accommodate. The speed of sound, for low densities ($n \lesssim n_0$), is believed to be much smaller than the causal limit; $c^2_s \ll 1$. At asymptotically high densities $n \gg n_s$ where conformal limit is restored, $c^2_s$ approaches $1/3$ from below \citep{Altiparmak_2022, McLerran_cs, Leonhardt_2020, PAL2022122464, Brandes_2023, Ecker_2022}. The nature of $c^2_s$ at intermediate densities is uncertain, as EoSs are still vague. However, there is a strong belief that $c^2_s$ is most likely to cross the conformal limit i.e. ($c^2_s > \frac{1}{3}$) to satisfy the maximum mass bound of $M > 2M_\odot$ \citep{Bedaque_2015, Hoyos_2016, Moustakidis_2017, Altiparmak_2022, Nath_2023}. The speed of sound, being the derivative, is highly sensitive to changes in $P$ with $\epsilon$. This sensitivity enables the detection of any discontinuities in the EoS and can indicate potential phase transitions.

Another quantity that sparked interest among the community is the so-called trace anomaly. It is the normalized trace of the star's Energy Momentum tensor; $T^{\mu \nu}$, defined as:
$\Delta = \frac{1}{3}(\frac{g_{\mu \nu}T^{\mu \nu}}{\epsilon})$
The interest in this quantity is due to the fact that it provides a simple measure to test conformal symmetry in the interior of the star and hence at densities above $n_0$. For a perfect fluid, {$\Delta = \frac{1}{3} - \frac{p}{\epsilon}$}. Causality and thermodynamic stability asserts the allowed range for $\Delta$ to be $\frac{-2}{3} \leq \Delta \leq \frac{1}{3}$. At very high densities where conformal symmetry is restored, we have $\Delta = 0$. At finite densities, its behaviour is still unknown; however, it was conjectured $\Delta \geq 0$ in neutron star interiors by Fujimoto et al.\cite{Fujimoto_2022}. It was shown \cite{Ecker_2022} that the interiors of massive stars can have values of $\Delta$ to be either positive, zero or negative. 

Although pulsar observation and theoretical understanding have enabled us to confine the matter properties to quite a large extent, it is still incomplete. One of the main obstructions is the large uncertainty in the radius measurement. We have only two pulsars whose radius is measured with an uncertainty of 2 km or more. Neutron Star Interior Composition Explorer (NICER), a soft X-ray telescope installed on the International Space Station, was developed to estimate masses and radii of NSs using pulse-profile modelling of nearby millisecond pulsars (MSPs). Pulse-profile modelling technique uses general relativistic effects on thermal emission from hot regions on the stellar surface \citep{Bogdanov_2019}. Two independent groups reported the radius of PSR J0030+0451 to be $12.71^{+1.14}_{-1.19}$ km by \citet{Riley_2019} and $13.02^{+1.24}_{1.06}$ km by \citet{Miller_2019}. Based on fits of rotating hot spot patterns from NICER as well as Xray - Multi Mirror (XMM Newton) data, the radius of PSR J0740 + 6620 was reported to be $13.71^{+1.30}_{1.50}$km by \citet{Miller_2021}. Accompanied by an additional 1.1 Ms exposure, the updated radius for PSR J0740 + 6620 has been reported to be $12.92^{+2.09}_{1.13}$km \citep{dittmann2024precisemeasurementradiuspsr}. 

Therefore, we need radius measurement of more pulsars with different masses and more accuracy. In this work, we study how future radius measurements of more pulsars with different mass windows and similar accuracy help us constrain the EoS further. The study addresses the effect observations have on constraining EoS. Furthermore, we have shown how EoS gets constrained in the $\epsilon-P$ space for a given radius window and analyzed the stiffness such EoS are expected to possess by studying the speed of sound ($c^2_s$). 

The paper is arranged as follows: Section II describes how the EoS ensemble is constructed and the different constraints imposed on it. Section III presents our results, and section IV summarises our results and draws a conclusion.

\section*{Formalism}
To study the effect of future mass and radius constraints, one must explore and define the entire EoS regime allowed with present astrophysical and thermodynamic bounds. Once the present bounds are well defined, one can observe the constraints imposed upon it with future mass and radius observations. To perform this, we need to build an ensemble of EoS from thermodynamic consistency and the bound on the EoS at low and asymptotic high density.  

Physics below saturation density $n_0$ can be well described by nuclear theory. EoS for matter below $n_0$ has been worked upon, and models such as Baym-Pethick-Sutherland (BPS), and Chiral Effective Theory (CET) describe the state of matter below saturation density. The ensemble of EoS used in this study has been constructed in segments. Baym–Pethick–Sutherland (BPS) tabulated equation of state is used till density $\sim 0.57 n_0$. Between the range of $0.57n_0 \lesssim n \lesssim 1.1n_0$, we construct monotropes of the form $p(n) = K{n^\Gamma}$, where we set the value of K on the extreme limits of the BPS EoS. The value of $\Gamma$ is sampled between $\left[1.77, 3.23\right]$ to span the CET band, ensuring that pressure is strictly in the range of stiff and soft EoS of Hebeler et al. \citep{Hebeler_2013}. For the densities $1.1n_0 \lesssim n \lesssim 80n_0$, the ensemble of EoS has been constructed using the sound-speed interpolation method introduced by \citet{Annala2020}

To employ this method, the speed of sound ($c_s^2$) is parametrized as a function of chemical potential $\mu$. The number density is then expressed as:\par
\begin{equation}
n(\mu) = n_{\scriptscriptstyle CET}\exp\left[\int_{\mu_{CET}}^{\mu} \frac{d\mu'}{\mu' c_s^2(\mu')}\right]    
\end{equation}
where $n_{\scriptscriptstyle CET}$ = 1.1$n_0$; $\mu_{\scriptscriptstyle CET}$ is the chemical potential corresponding to $n_{\scriptscriptstyle CET}$.\par
Pressure is given by:
\begin{equation}
p(\mu) = p_{\scriptscriptstyle CET} \! + n_{\scriptscriptstyle CET}\int_{\mu_{CET}}^{\mu}d\mu' \exp\left[\int_{\mu_{CET}}^{\mu'}\frac{d\mu''}{\mu''c_s^2(\mu'')}\right]
\end{equation}
where $p_{\scriptscriptstyle CET} \equiv p(\mu_{\scriptscriptstyle CET})$. The number of linear segments (N) for the speed of sound and take N pairs sequence of $\left\{\mu_{i},c_{s,i}^{2}\right\}_{i=1}^{N}$ is fixed. The values at the boundaries are matched such that $\mu_{1} = \mu_{CET}$ and $\mu_{N} = 2.6 GeV$ to make sure that our EoSs are consistent with pQCD result for cold quark matter in beta equilibrium. A piecewise-linear function is employed to connect the points $\left\{\mu_{i},c_{s,i}^{2}\right\}_{i=1}^{N}$ which simply gives us:\par
\begin{equation}
c_s^2(\mu) = \frac{(\mu_{i+1} - \mu)c_{s,i}^2 + (\mu - \mu_{i})c_{s,i+1}^2}{\mu_{i+1} - \mu_i}
\end{equation}
It is to be noted that instead of sampling $c_{s,i}^2 \in [0,1]$, we rather choose $c_{s,max}^2$ from $[0,1]$ and then uniformly sample $\mu_{i} \in [\mu_{1},\mu_{N+1}]$ and $c_{s,i}^2 \in [0,c_{s,max}^2]$. Finally, we numerically integrate for pressure.


\section*{Results}

Once EoSs have been constructed, we solve TOV equations numerically to obtain the mass-radius sequence of non-rotating stellar models. With the mass-radius relations in hand, we can begin employing constraints to it. First, we reject those EoSs for which $M_{\scriptscriptstyle TOV}< 2M_\odot$. Here, $M_{\scriptscriptstyle TOV}$ is the maximum mass attainable for a given EoS. Next, we impose the measurements of PSRJ0030 + 0451: $M_{NS} = 1.44_{-0.14}^{+0.15}M_\odot, R = 13.02^{+1.24}_{-1.06} km$ as given in \citet{Miller_2019} as well as of PSRJ0740 + 6620: $M_{NS} = 2.08 \pm 0.07 M_\odot, R = 13^{+2.6}_{-1.5} km$ as given in \citet{Miller_2021}.
\begin{figure*}[t]    
    \centering
    \includegraphics[height = 6cm,width=.49\linewidth]{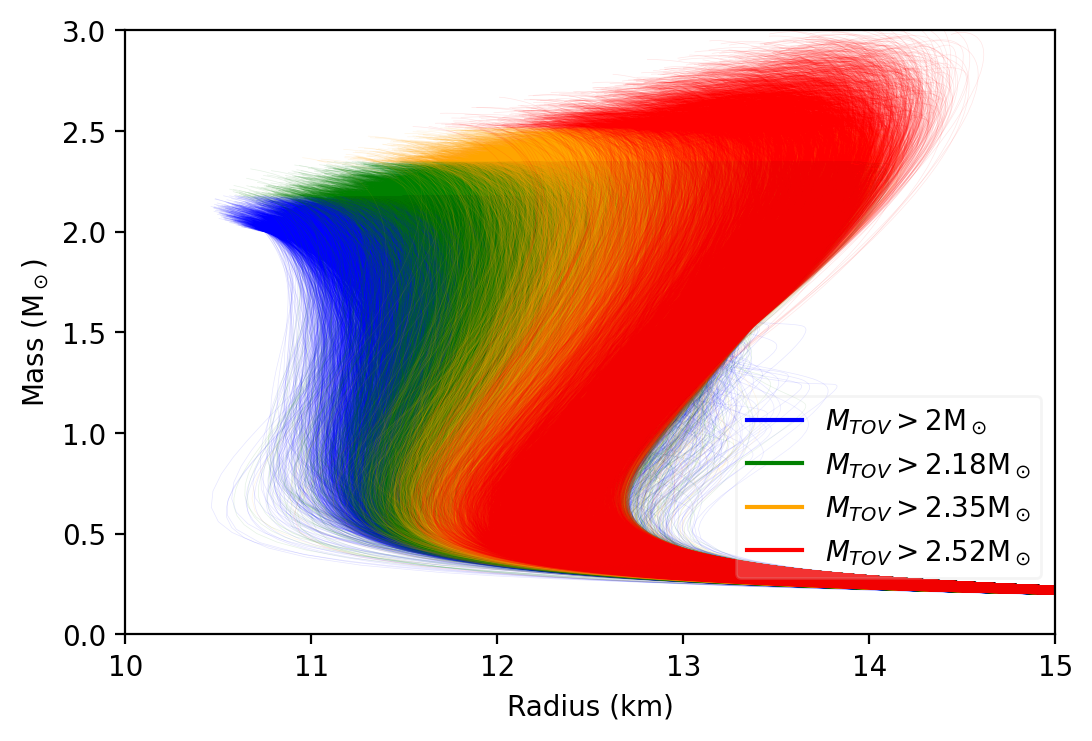}
    \includegraphics[height = 6cm,width=.49\linewidth]{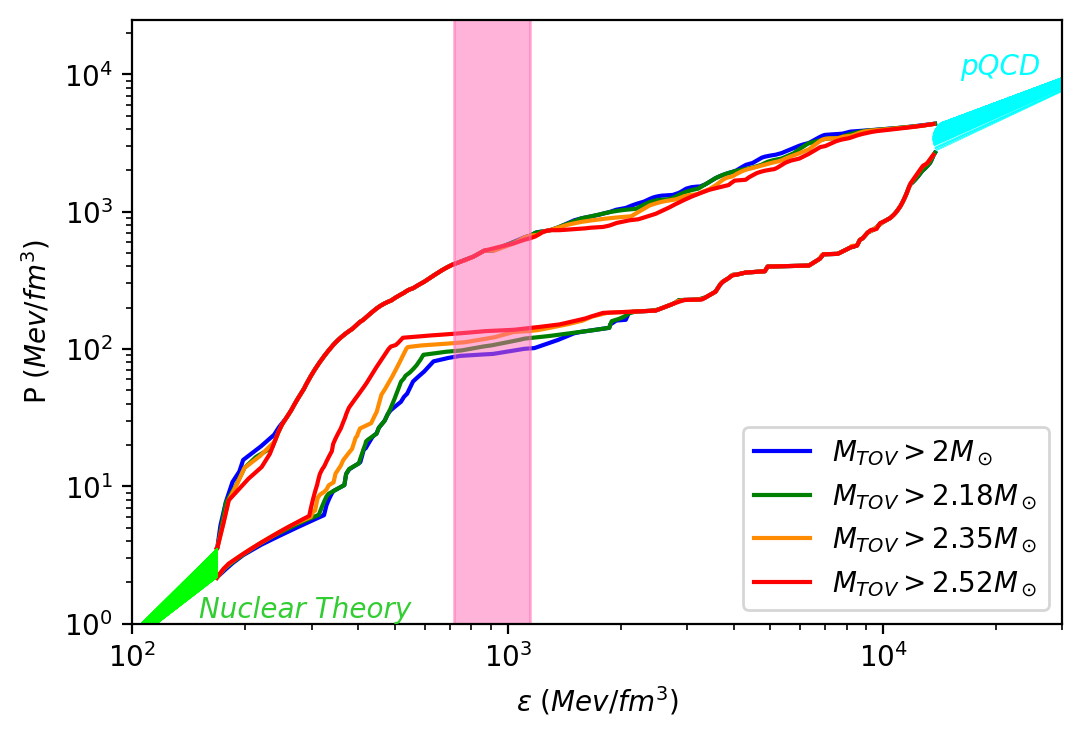}
    \includegraphics[height = 6cm,width=.49\linewidth]{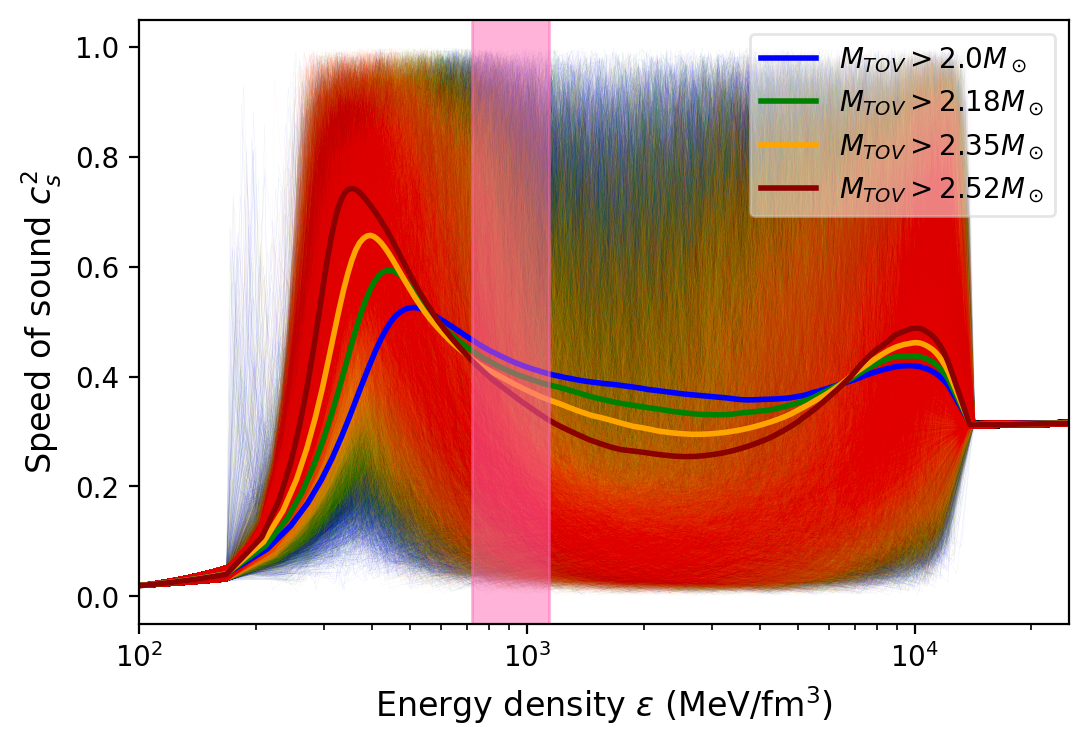}
    \includegraphics[height = 6cm,width=.49\linewidth]{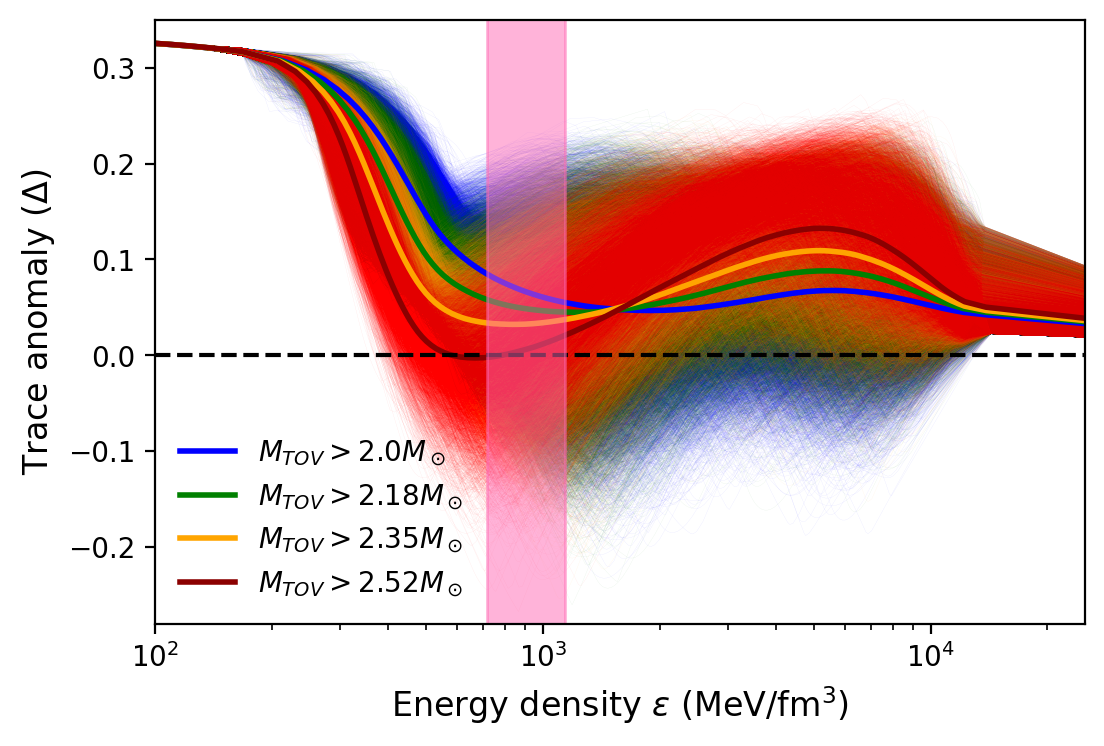}
    \caption{\textit{Top left}: M-R curves of the ensemble of EoS constructed. \textit{Top right}: Contours of EoSs for various mass constraints as indicated. The green and cyan band shows the uncertainty in EoSs for nuclear theory and pQCD limits, respectively. \textit{Bottom left}: speed of sound as a function of energy density for the mass constraints and their respective means (in solid coloured lines). \textit{Bottom right}: Trace anomaly as a function of energy density for the mass constraints and their respective means (in solid coloured lines). The pink strip of the plots indicates the 95\% confidence for the densities of $M_{TOV}$ achieved by our EoSs.}
    \label{fig:example}    
\end{figure*}
We have analyzed various mass values of non-rotating NSs by parametrizing the radius window and observing how the EoS region gets constrained. We have also shown the speed of sound($c_s^2$) trends and trace anomaly ($\Delta$) of EoSs by plotting their respective means.

We present the M-R relation of our constructed EoS in Fig \hyperref[fig:example]{\ref*{fig:example}}, top Left. As one moves towards higher masses, the MR curve segregation narrows. The minimum cutoff mass for the analysis done in this work is $2$ solar mass.

The EoSs contours corresponding to different mass segments are shown in Fig \hyperref[fig:example]{\ref*{fig:example}}, top Right. It shows how the EoSs corresponding to different values of $M_{TOV}$ are constrained in the $\epsilon-P$ space. One needs both the EoS ($\epsilon-P$) and the speed of sound plot to understand the behaviour of the EoS completely. For the minimum mass range, the EoS plot is the most wide. Indicating that both stiff and soft EoS are more or less evenly distributed. As we go to a higher mass range, although the upper limit of the EoS contour plot remains the same, the lower limit rises, indicating the favouring of stiffer and stiffer EoS. Stiffer EoS can produce higher $M_{TOV}$ stars. The speed of the sound curve also conveys an exciting result. As the mass range gets higher, the average speed of sound increases and increases more rapidly (the peak appears at a lower density). 

In the EoS plot, there is a slight kink at lower densities in the upper limit of the contour plot. This is because few EoS (supporting relatively low-mass stars) pass through the given region. It is also evident in the speed of the sound curve as few EoS in the low mass range attain their peak very early. However, they are statistically insignificant as they do not influence the average speed of sound much. Beyond the peak value, the mean value of $c_{s}^2$ for the heavier mass range is minimum, whereas for the min $M_{TOV}$ range, the mean value is maximum. The pink strip denotes the 95\%  confidence region where $M_{TOV}$ occurs. Beyond the given region, it is just the extrapolation of the curves maintaining the thermodynamical stability, causality and asymptotic pQCD bound.

Trace anomaly $(\Delta)$ for the various mass segments have been shown in Fig \hyperref[fig:example]{\ref*{fig:example}}, Bottom right. The mean of the trace anomaly for most massive stars with maximum mass segment falls off most rapidly, attaining a minimum of around zero and increasing again, having a maximum at very high density and asymptotically going towards zero. As the $M_{TOV}$ mass range decreases, the curve follows a similar trend; however, they are much flatter. Therefore, there is a crossing of the curves beyond the 95\%  confidence region of $M_{TOV}$.
The zero value of $\Delta$ is achieved by only the mean of $M_{TOV} \geq 2.52 M_{\odot}$. The speed of sound is related to the trace anomaly via $c_s^2 = (\frac{1}{3} - \Delta) - \frac{d\Delta}{d ln\epsilon}$ \citep{Fujimoto_2022}. In the conformal limit $\Delta \xrightarrow{}0$ and $\frac{d\Delta}{dln\epsilon} \xrightarrow{} 0$, which renders $c_s^2 \xrightarrow{} \frac{1}{3}$. Although trace anomaly infers about the EoS, once the pressure, energy density and speed of sound are analyzed, it does not provide much additional information. Therefore, we do not explicitly analyze them further in the text.

\begin{figure*}[t]   
    \includegraphics[height = 4.5cm,width=.49\linewidth]{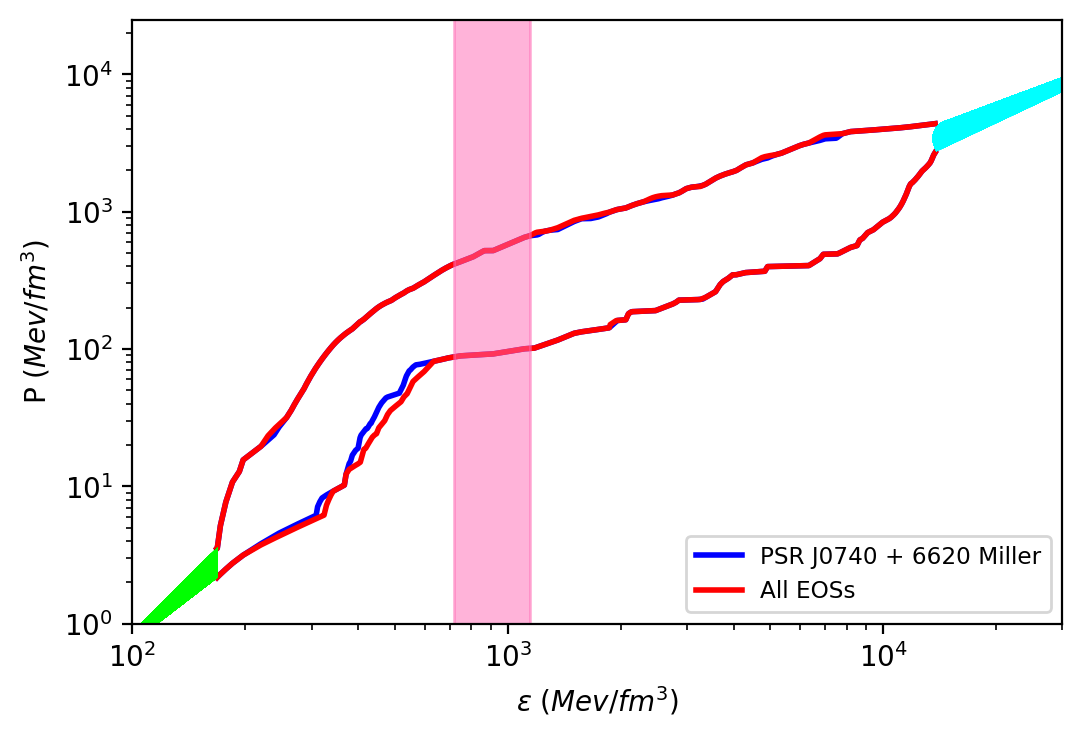}
    \includegraphics[height = 4.5cm,width=.49\linewidth]{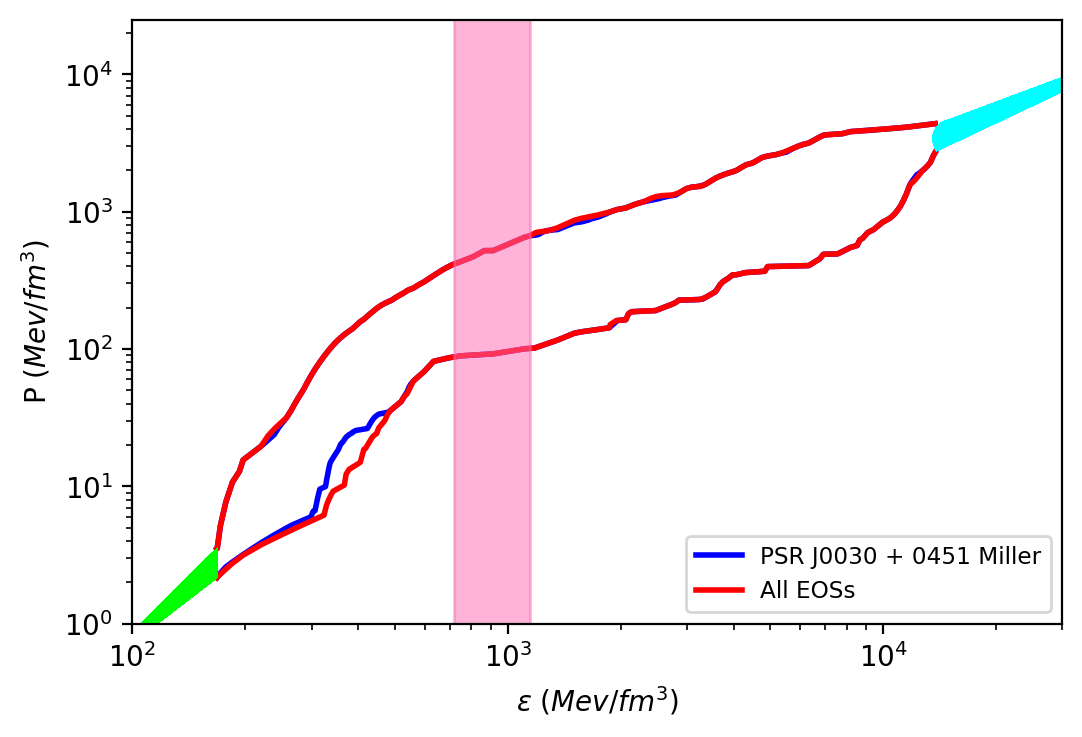}
    \caption{\textit{Left}: Contour of all EoS (red) with those that satisfy the constraints of PSR J0740 + 6620 (blue). \textit{Right}: Contour of all EoS (red) with those that satisfy the constraints of PSR J0030 + 0451 (blue).}
    \label{2}  
\end{figure*}
The precise measurement of mass has helped constrain the EoS for matter at the core of the NSs, and the simultaneous measurement of mass and radius by NICER has also refined the EoS limits. In this work, we employ the measurements by NICER for PSR J0740 + 6620: $M_{NS} = 2.08 \pm 0.07 M_\odot, R = 13^{+2.6}_{-1.5} km$ \cite{Miller_2019} and PSR J0030 + 0451: $M_{NS} = 1.44_{-0.14}^{+0.15}M_\odot, R = 13.02^{+1.24}_{-1.06} km$\cite{Miller_2021}.

We have also traced out the set of EoSs that satisfy both the pulsar constraints simultaneously and presented their resulting contours in Fig \hyperref[2]{\ref*{2}}. 

\begin{figure*}[t]
    \includegraphics[height = 5cm, width =0.49\linewidth]{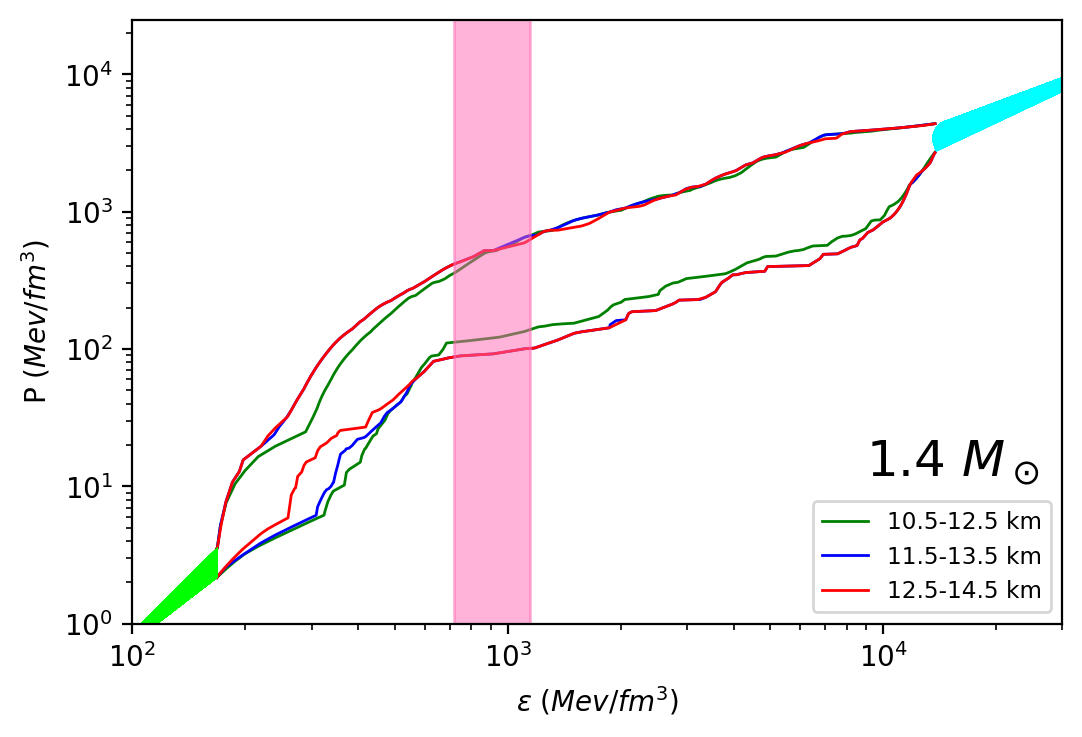}
    \includegraphics[height = 5cm, width =0.49\linewidth]{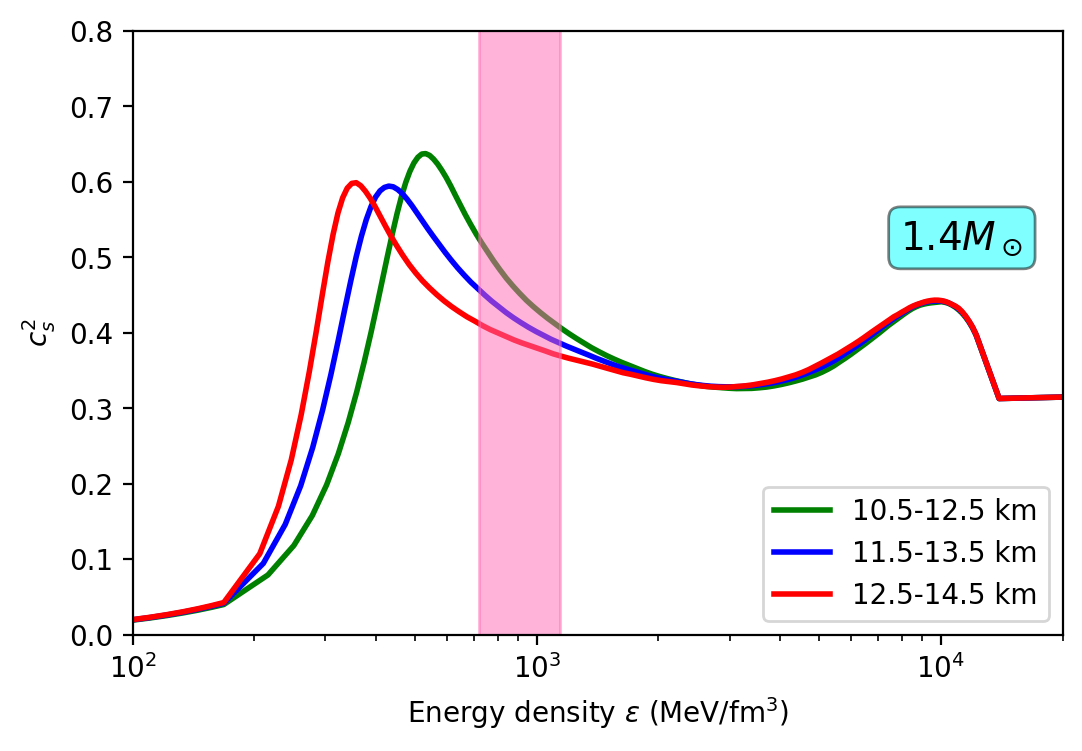}
    \includegraphics[height = 5cm, width =0.49\linewidth]{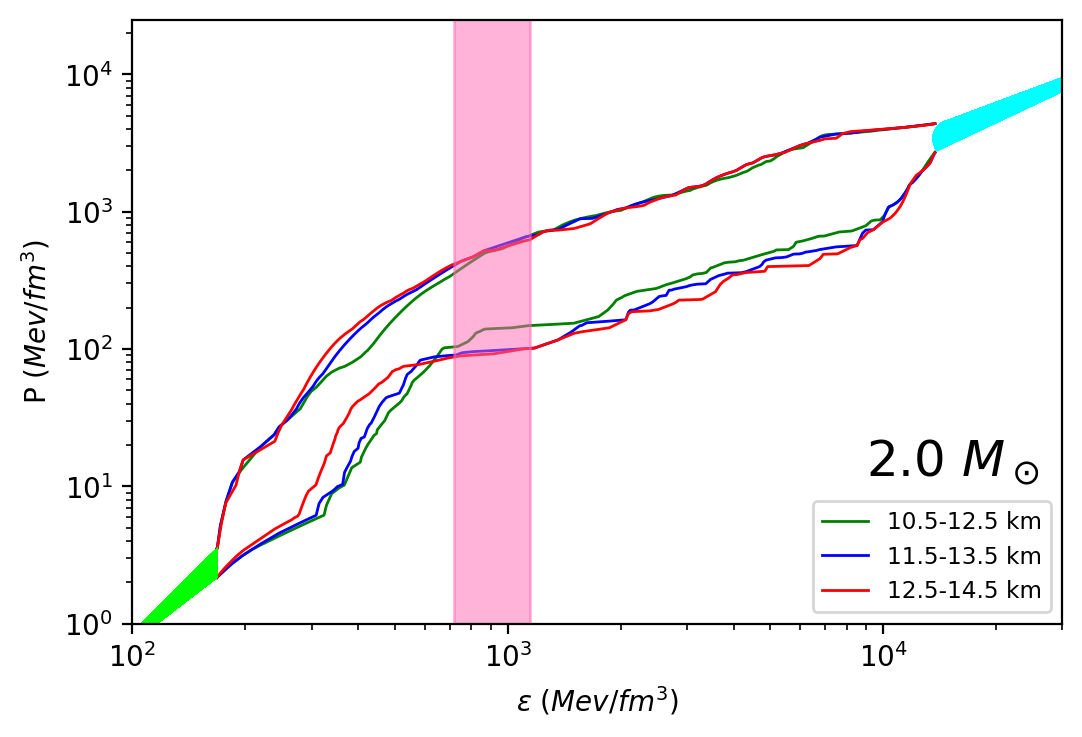}
    \includegraphics[height = 5cm, width =0.49\linewidth]{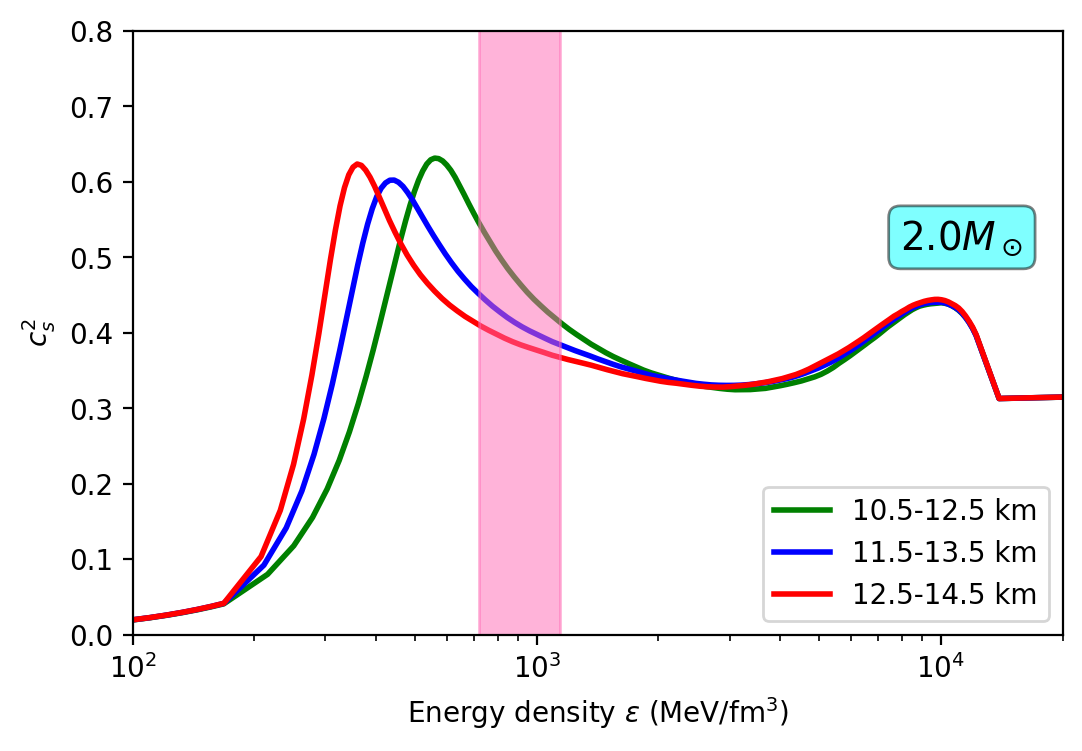}
    \includegraphics[height = 5cm, width =0.49\linewidth]{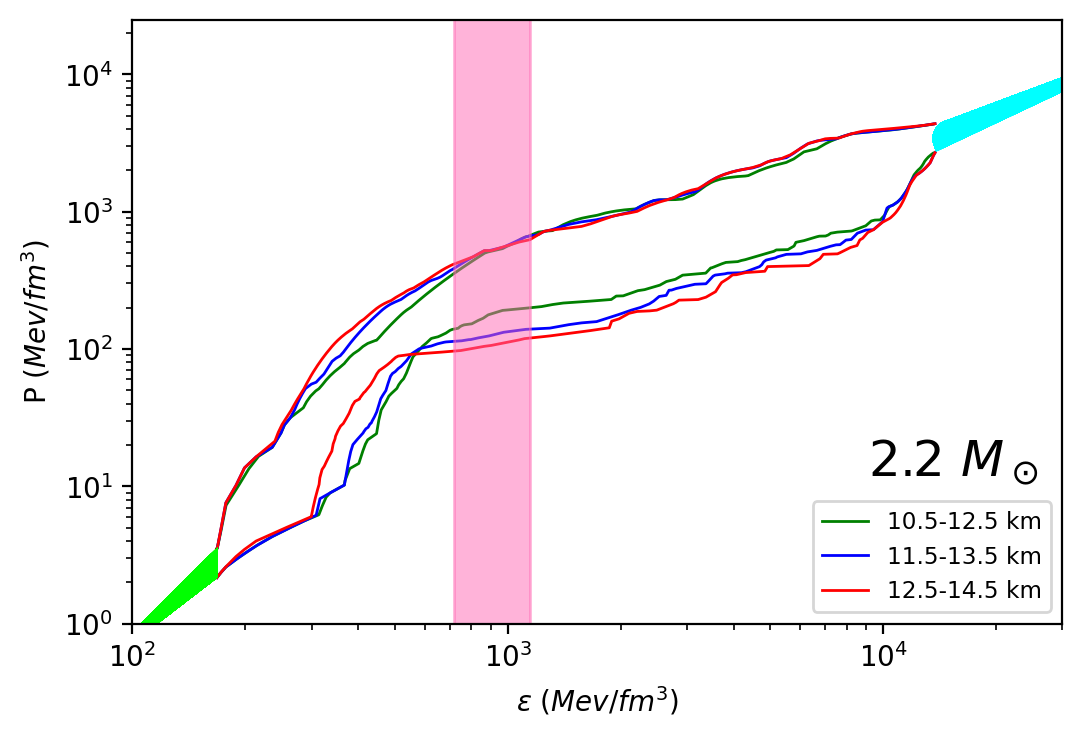}
    \includegraphics[height = 5cm, width =0.49\linewidth]{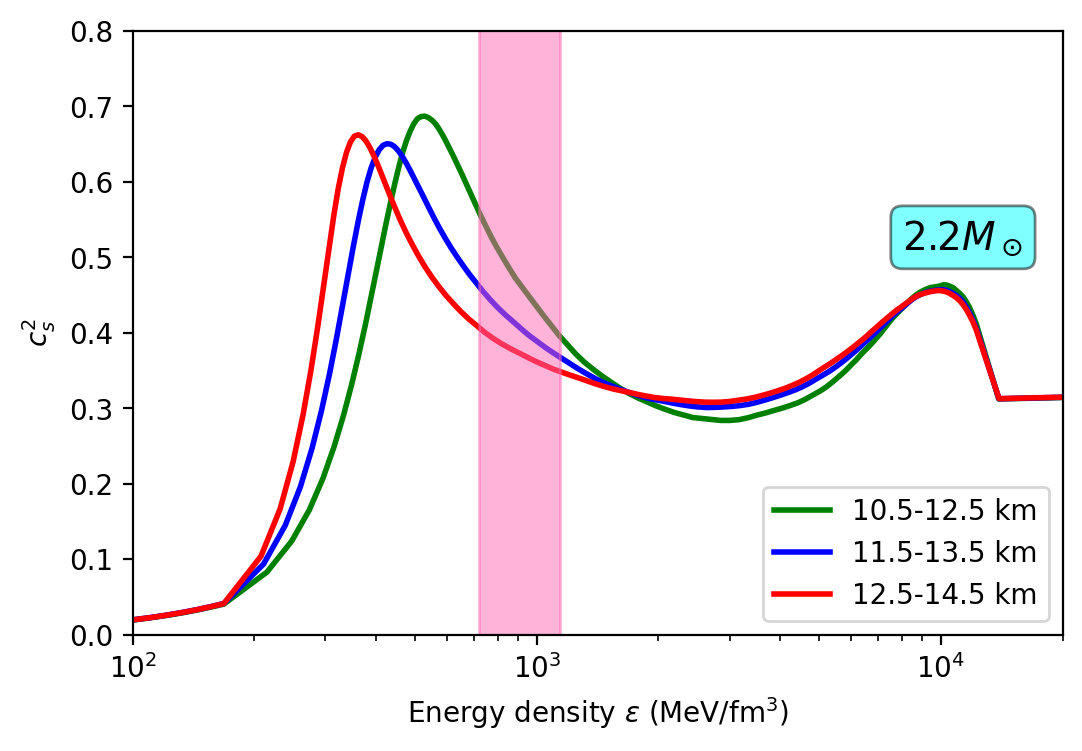}
    \caption{EoS contour plots for the radius range of 10.5-12.5 km (in green), 11.5-13.5 km (in blue), and 12.5-14.5 km (in red) have been shown for various masses in the left panel. The corresponding speed of sound ($c_s^2$) plots with energy density have been shown in the right panel.}
    \label{3}
\end{figure*}

Comparing both the constraints, we see that PSR J0030 + 0451 allows for stiffer EoS (right, Fig \hyperref[2]{\ref*{2}}) as compared to PSR J0740 + 6620 (left, Fig \hyperref[2]{\ref*{2}}). The lower half of the contour plot is reduced to a larger extent by the former than by the later NICER measurement, whereas the upper boundary remains the same. Having a stricter radius constraint for the data of PSR J0030+0451 allowed for filtering out more EoSs compared to PSR J0740+6620. 

The mass and radius measurements significantly affect the EoS governing the property of matter at NS cores. However, there are only a few precise mass and radius measurements. Although the effect of measurement of massive stars is evident, as discussed previously in the paper, the effect of radius measurement is not entirely clear as we only have two measurements. Therefore, we employ two methods to study the effect of radius measurement for future detectors and how they affect the matter properties of NSs.

\begin{figure*}[t]
    \includegraphics[height = 5cm, width =0.49\linewidth]{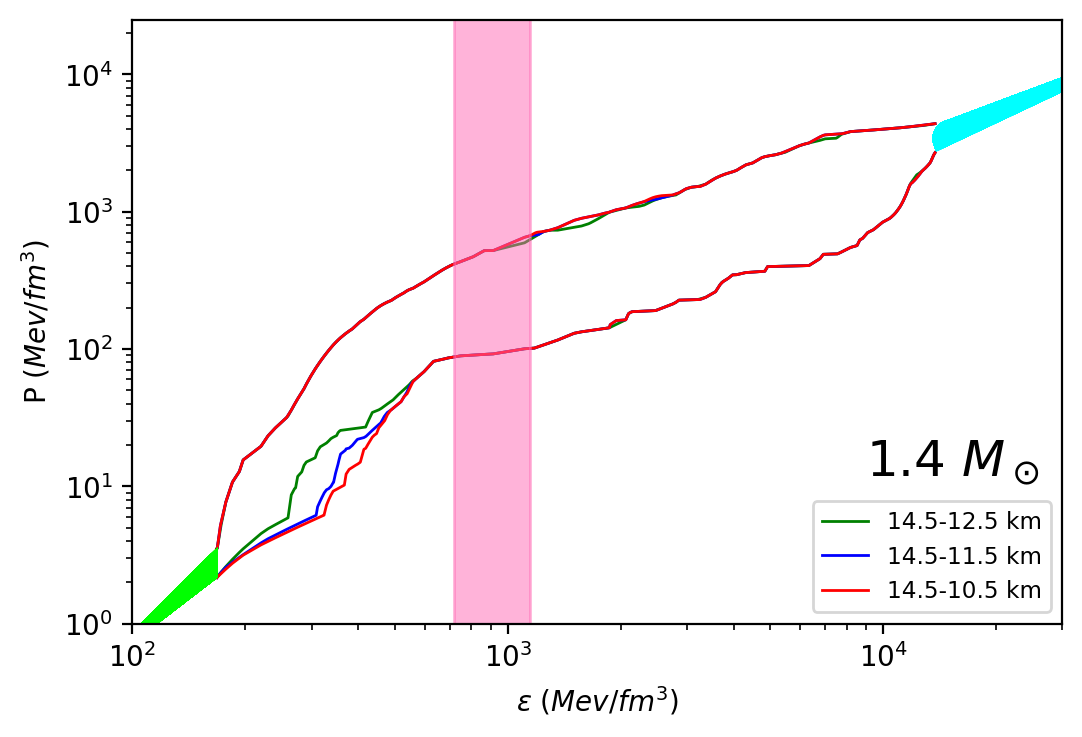}
    \includegraphics[height = 5cm, width =0.49\linewidth]{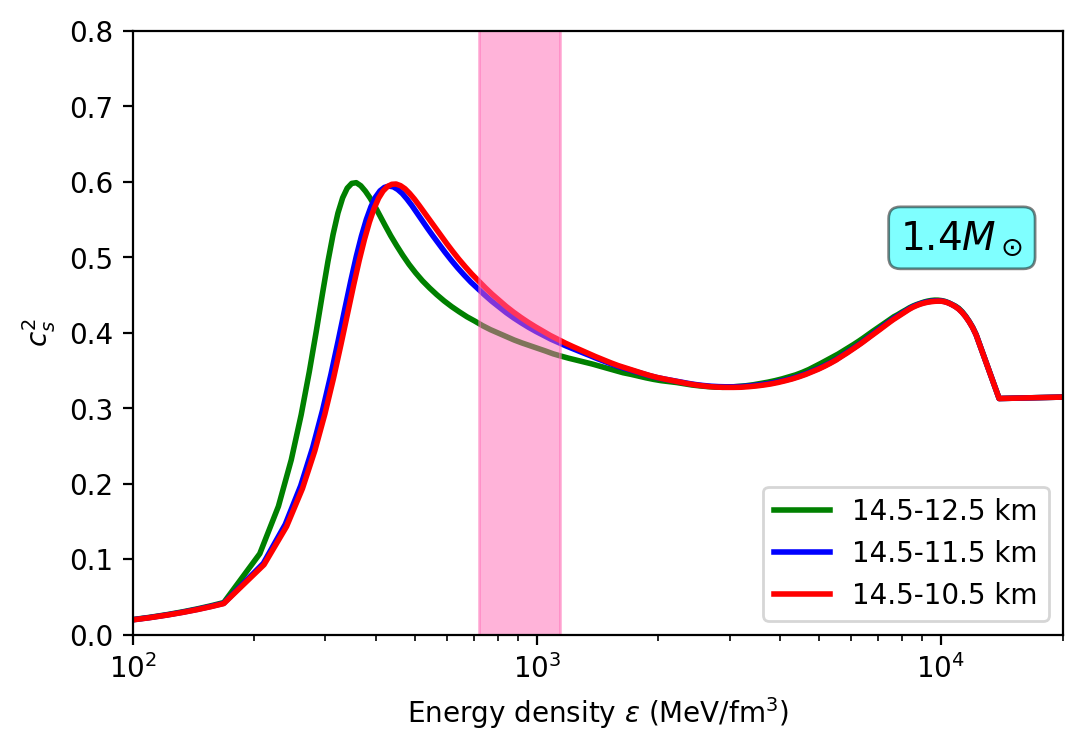}
    \includegraphics[height = 5cm, width =0.49\linewidth]{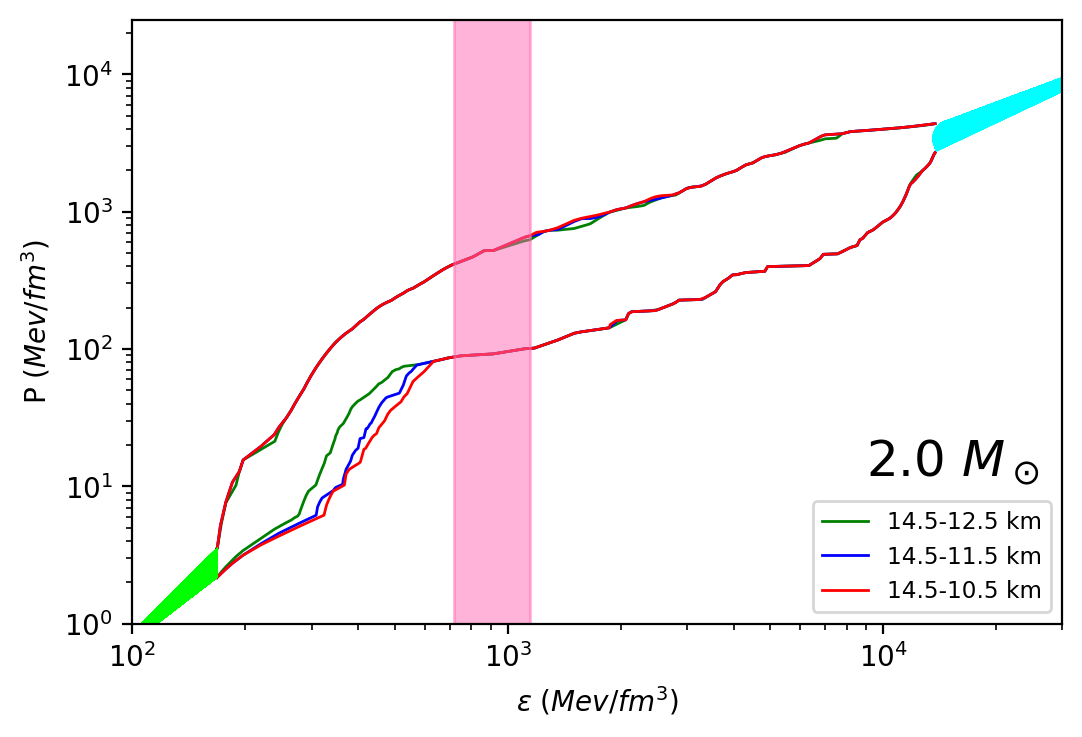}
    \includegraphics[height = 5cm, width =0.49\linewidth]{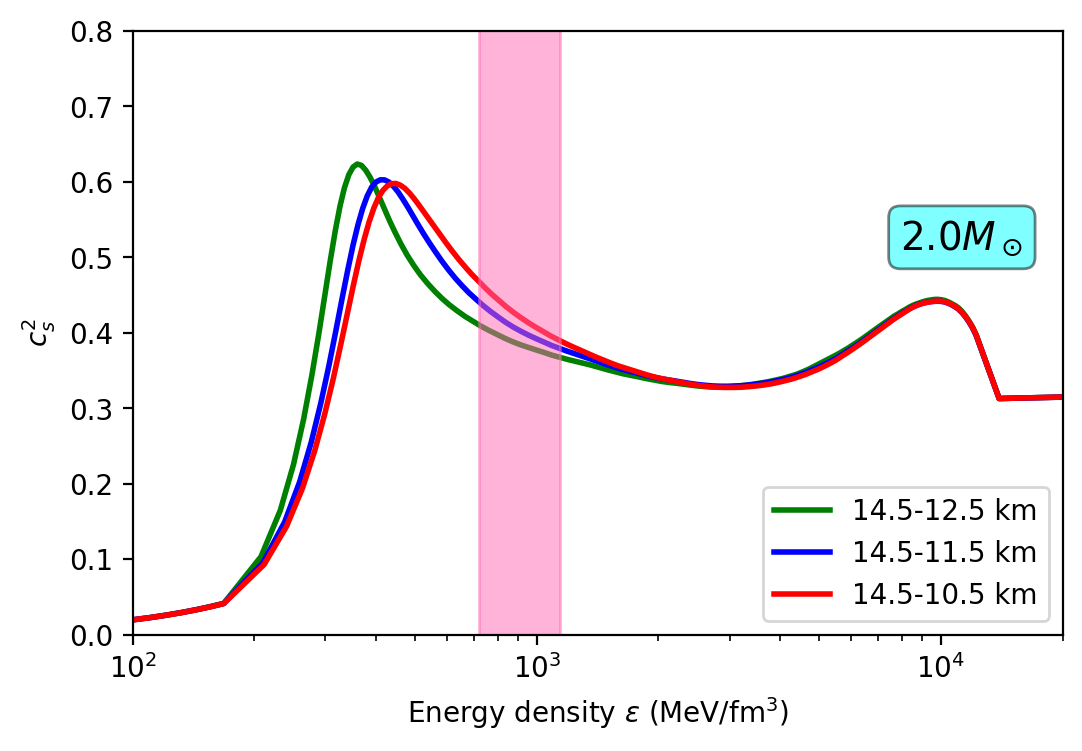}
    \includegraphics[height = 5cm, width =0.49\linewidth]{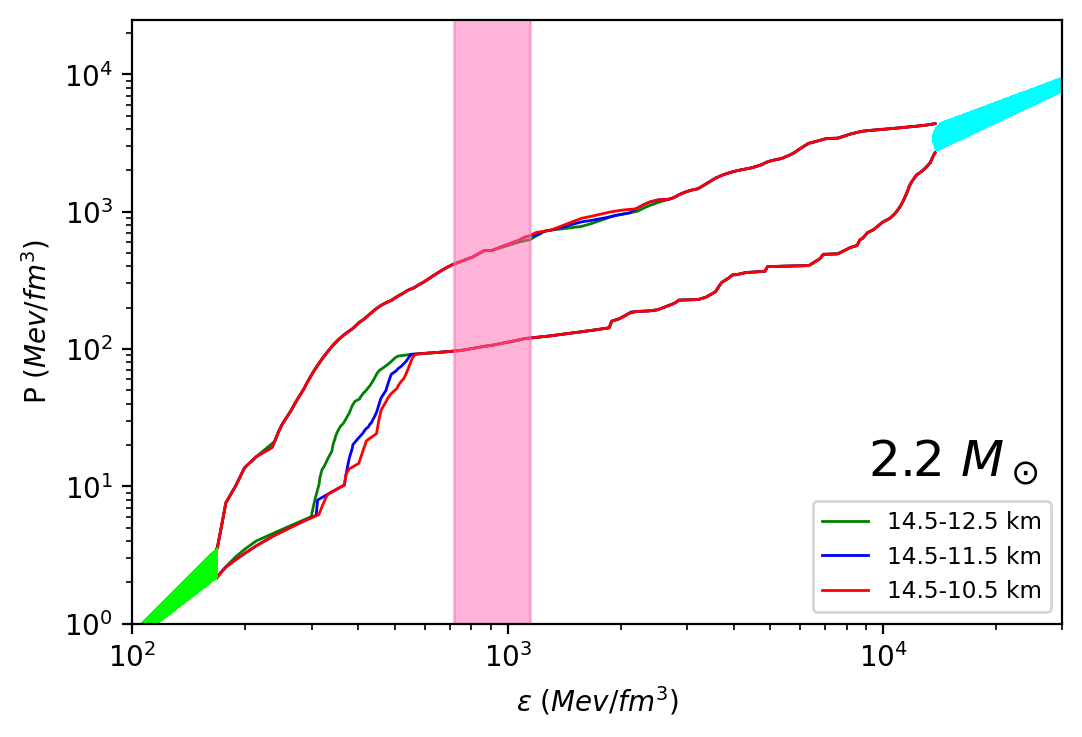}
    \includegraphics[height = 5cm, width =0.49\linewidth]{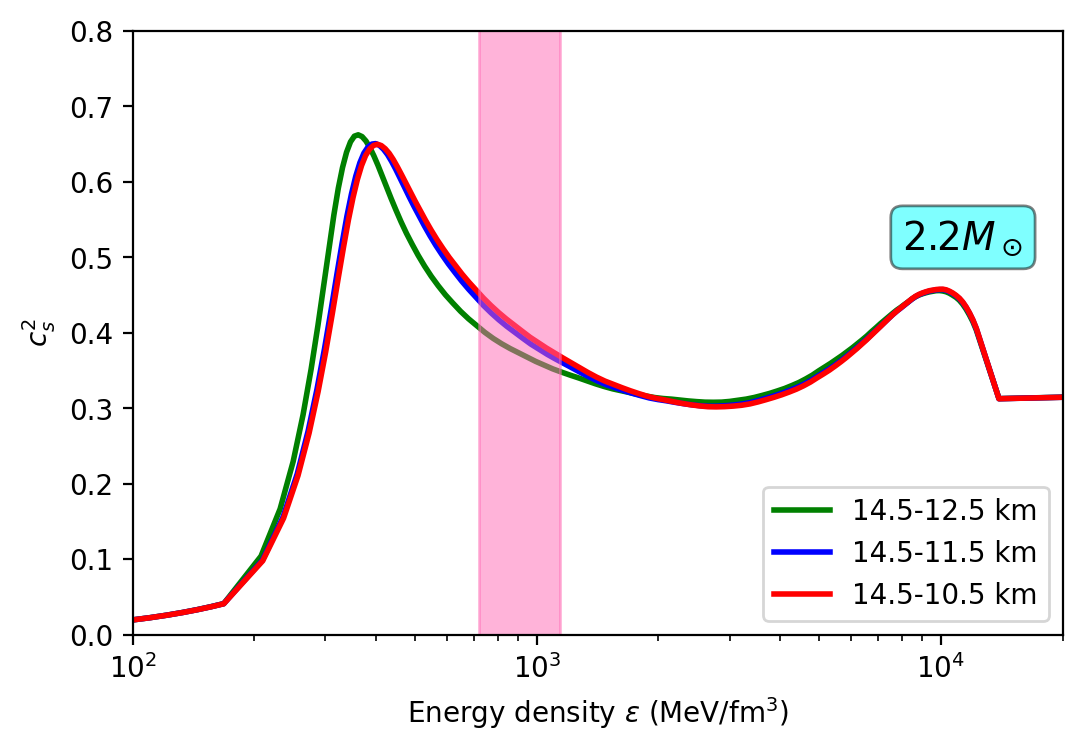}
    \caption{The contour plots of EoS with increasing radius window for various masses keeping the maximum radius fixed (left) with their corresponding plots for $c_s^2$ (right).}
    \label{4}
\end{figure*}

In the first method, assuming that the radius measurement is precise enough to a 2 km radius (for all masses), we select three radius windows: 10.5-12.5, 11.5-13.5 km, and 12.5-14.5 km. We check our result for 3 masses of NS 1.4 $M_{\odot}$, 2 $M_{\odot}$ and 2.2 $M_{\odot}$. To have a detailed analysis of our result, we plot both the contour and speed of sound graphs in Fig \hyperref[3]{\ref*{3}}. For 1.4 solar mass, we see that as the radius window shifts towards the higher side, the enclosed region of the contour curves shifts towards higher pressure values, indicating a stiffening of the EoS. This is expected as a higher radius indicates stiffer EoS for a given mass value. This is also clear from the corresponding speed of sound curve. As the radius window shifts towards a higher value, the mean of the $c_s^2$ becomes stiffer at smaller energy density values. However, after the peak, there is a fall in the speed of sound value, and the stiffness is reversed. As the enclosed region by the contour curves shifts entirely, both the peak value and the location of the peak of the mean of speed of sound changes.

\begin{figure*}
    \includegraphics[height = 5cm, width =0.49\linewidth]{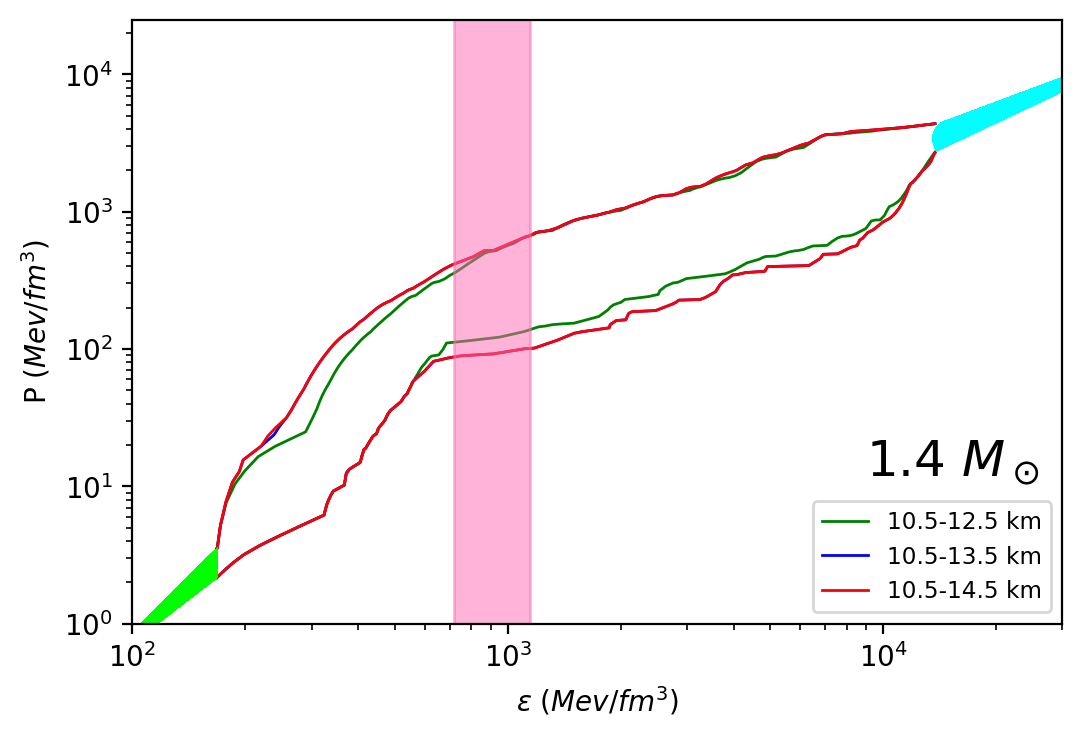}
    \includegraphics[height = 5cm, width =0.49\linewidth]{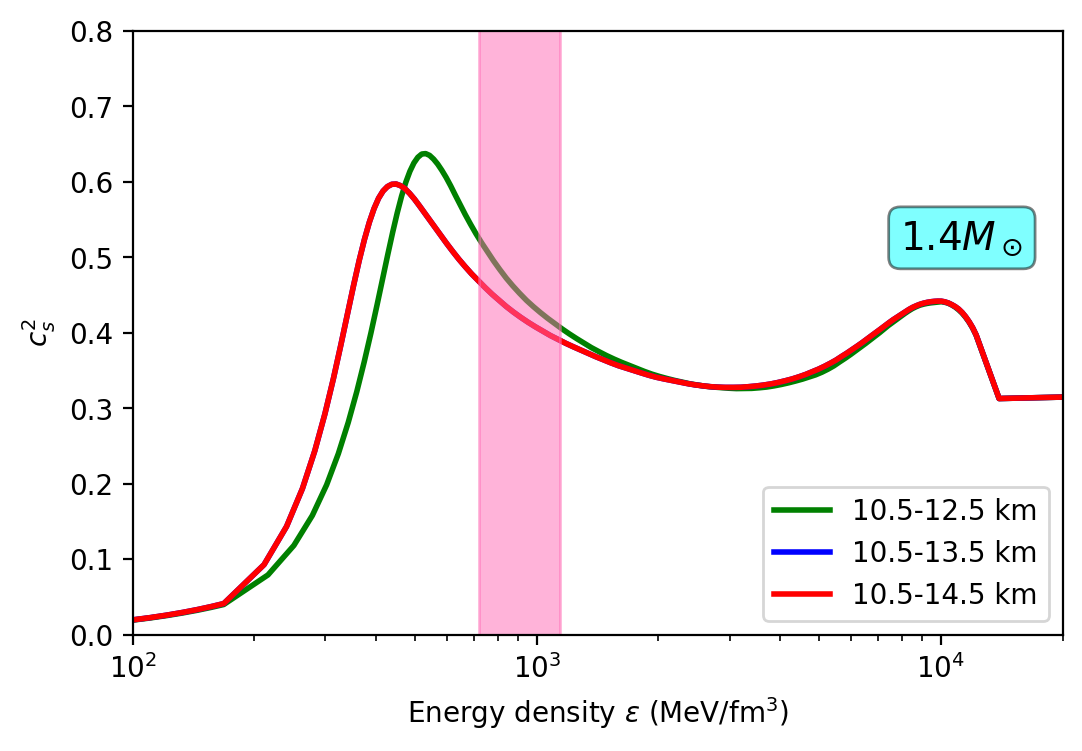}
    \includegraphics[height = 5cm, width =0.49\linewidth]{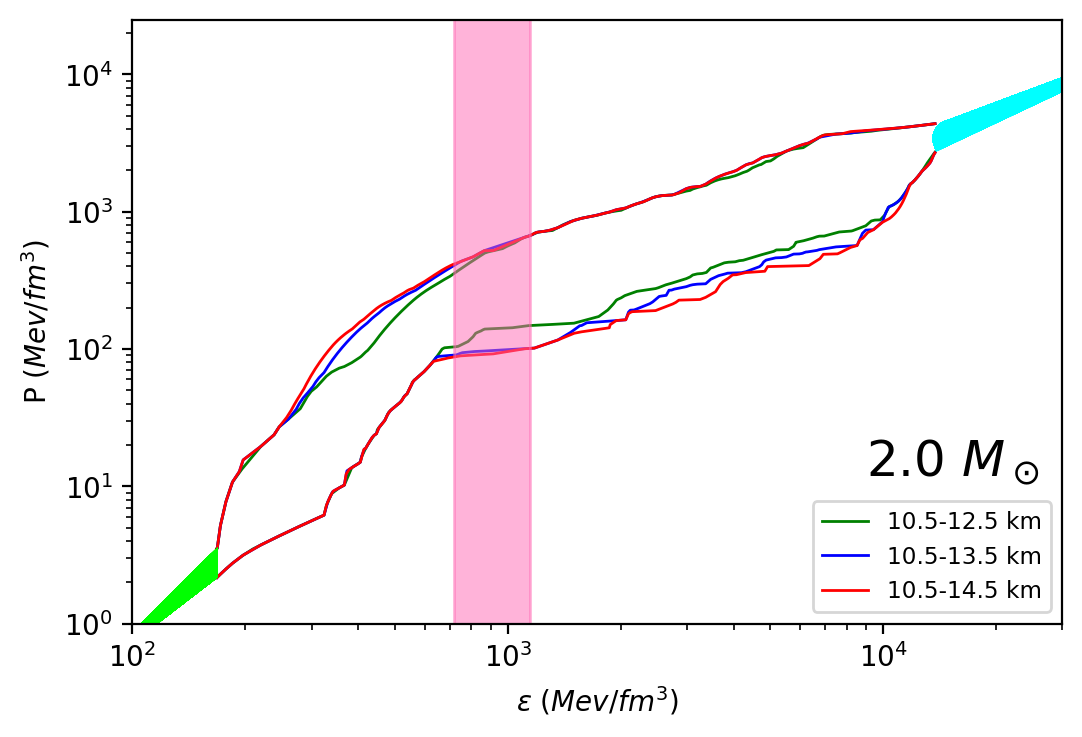}
    \includegraphics[height = 5cm, width =0.49\linewidth]{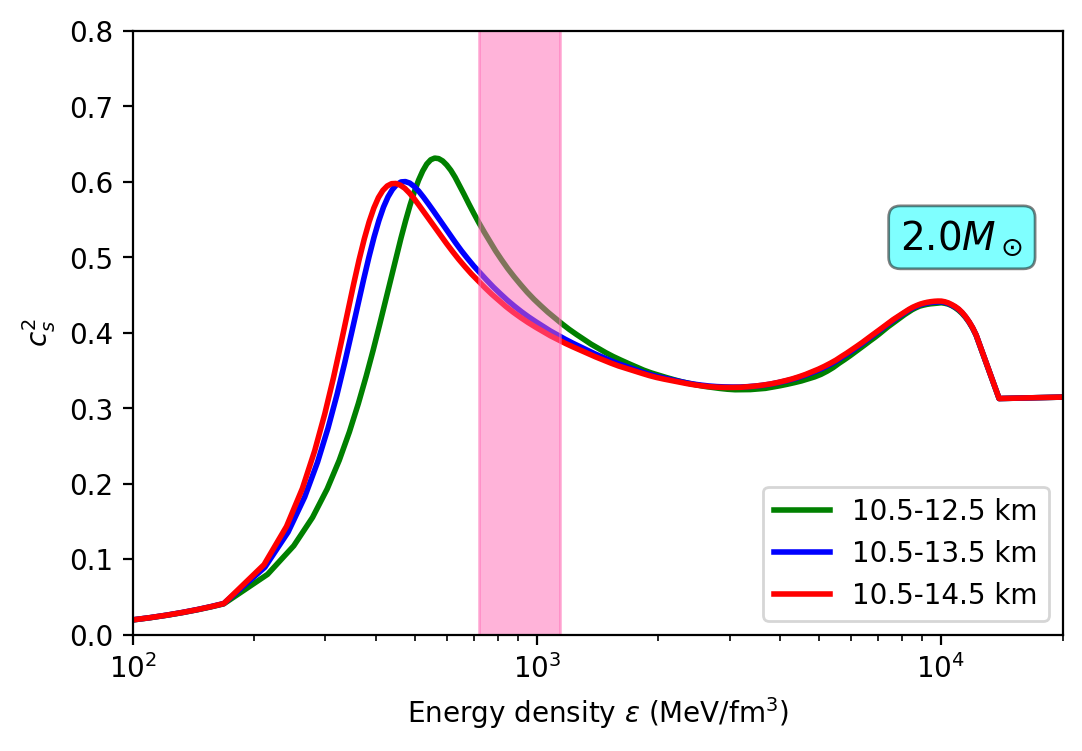}
    \includegraphics[height = 5cm, width =0.49\linewidth]{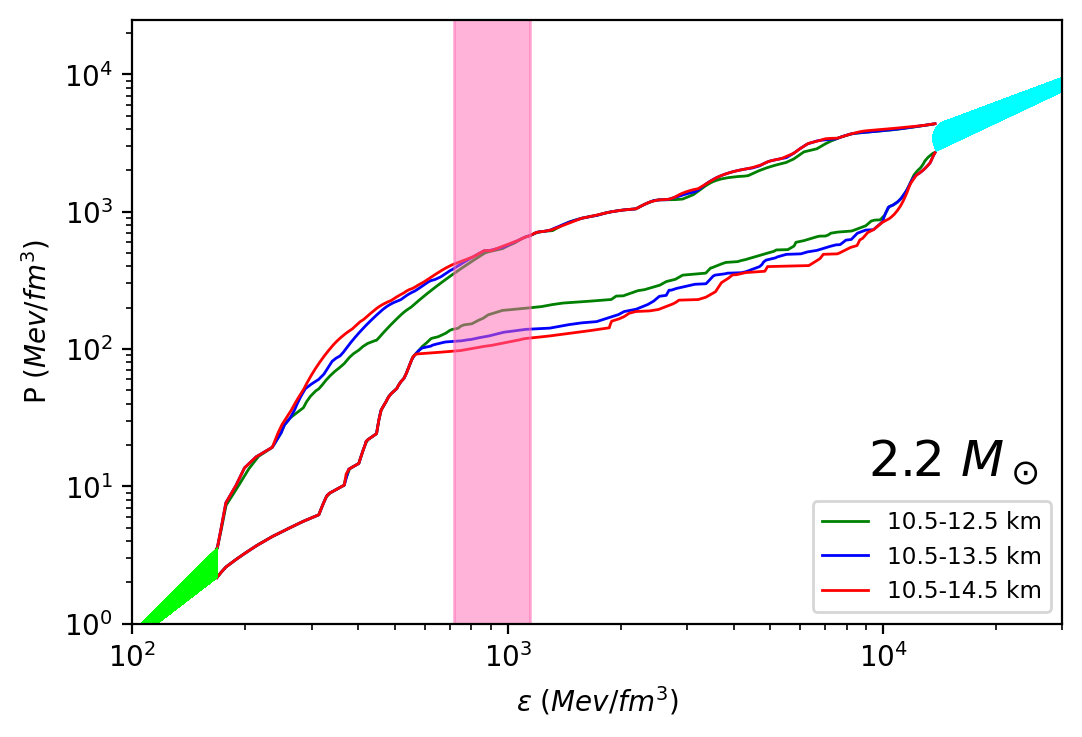}
    \includegraphics[height = 5cm, width =0.49\linewidth]{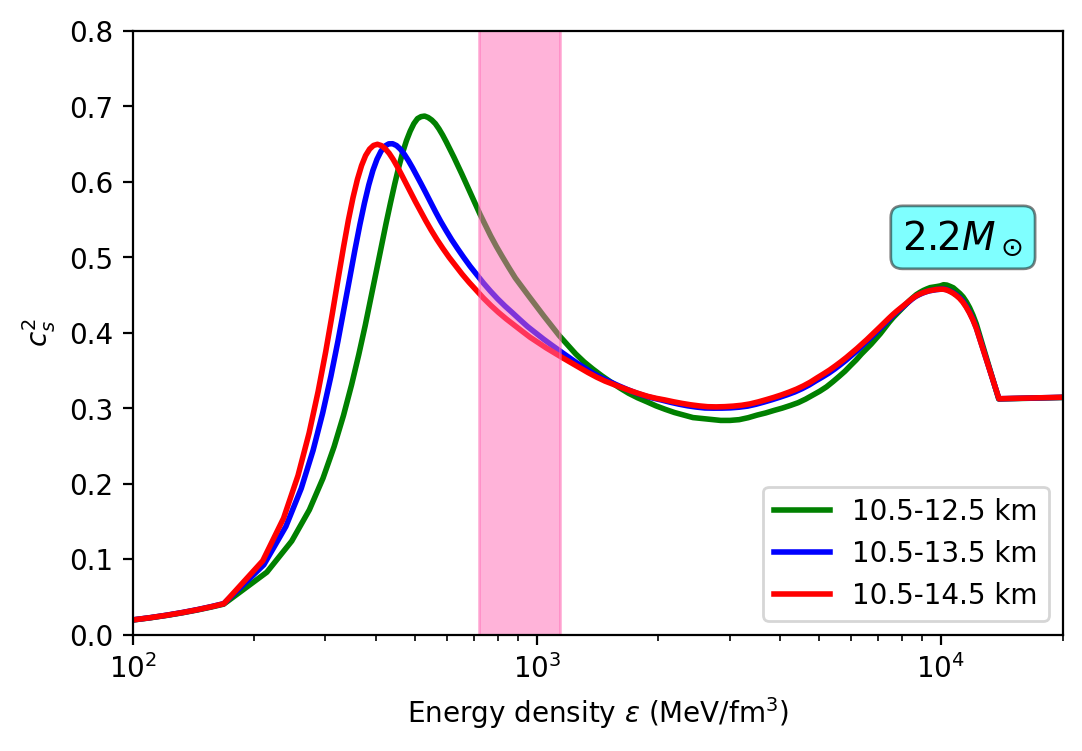}
    \caption{The contour plots of EoS with increasing radius window for various masses keeping the minimum radius fixed (left) with their corresponding plots for $c_s^2$ (right).}
    \label{5}
\end{figure*}

A similar pattern is also seen for the other two masses (2.0 $M_{\odot}$ and 2.2 $M_{\odot}$); however, the quantitative value changes. As the mass increases, the height of the peak increases slightly. Also, looking at the plots, we see a shift in the energy density value, where the peak of $c_s^2$ occurs. For the increasing value of the radius window, the value of energy density, where $c_s^2$ peaks, is found to be decreasing. The reversal of the nature of stiffness beyond the peak value indicated for $M_{TOV}$ stars, the larger stars have softer cores, while the smaller stars have stiffer cores.
It can also be seen that the equation of states producing larger stars attains their peak values at lower densities compared to those producing smaller stars. Such a trend is observed for all our mass values.

It is not always apparent that the radius precision is always 2 km for all mass values. It can vary with mass, detectors and the pulsar's distance from us. Therefore, a comparison with varying radial precision is done using the second method. It is done in two ways: one while keeping the maximum radius fixed and varying the radial precision, and another by keeping the minimum radius fixed and varying the radial precision. The maximum radial precision is kept at 2 km, and the minimum at 4 km (as almost all the M-R curves in this work lie between 10.5-14.5 km, a 4 km window). This is again done for three sets of masses, and we have compared the contours and speed of sound plots to have a detailed study.
Here, we try to observe how the region of $\epsilon-P$ space gets affected when we change the radius window for a given value of stellar mass. However, it is made sure every EoS produces $M_{TOV} > 2M_{\odot}$.

For the former case, where the maximum radius is kept fixed (14.5 km), we show our result in Fig \hyperref[4]{\ref*{4}}.
First, we analyze the 1.4 $M_{\odot}$ results. As the precision increases, the lower bound of the contour plot shrinks, indicating a stiffening of the EoS. This is also clear from the speed of sound plot. At lower energies, the speed of sound is maximum for the most precise radial value measurements and attains a peak at the earliest. For the lower precise measurements, the speed of sound is attained at higher energy densities indicating softer EoS. However, after the peak values, the stiffness ordering is reversed. The most precise measurement gives the least stiff stars (at higher densities). 
Again, the trend is similar for higher-mass stars (2 $M_{\odot}$ and 2.2 $M_{\odot}$). However, the peak maximum value increases slightly with the increase in mass.

Finally, we analyze the problem where the minimum radius is kept fixed ($10.5$ km), and the radial precision varies (Fig \hyperref[5]{\ref*{5}}). For 1.4 $M_{\odot}$, as the precision increases, the upper boundary of the contour curves shrinks, indicating a softening of the EoS. This is expected for a given mass; if the radius is low, it indicates a softer EoS. This is also clear from the speed of sound curve, where, for the least precise radial measurement, the mean of the speed of sound attains its maximum value as soon as possible. The more precise measurement attains a peak at relatively higher densities, indicating softer EoS. However, again, after the peaks, the stiffness nature of the curves is completely reversed. The trend remains more or less similar for higher-mass stars, only changing quantitatively.

\section{Summary And discussion}

 In this work, we have studied how mass and radius measurement can help constrain the EoS of matter inside an NS. As one measures higher and higher masses of NSs, the EoS gets constrained (primarily to 8-10 times) nuclear saturation density. This results in the stiffening of the EoS, which is revealed in the speed of the sound curve. For higher masses, the speed of sound not only reaches the maximum sound speed but also reaches the earliest. This aspect is also revealed in the trace anomaly plots. The trace anomaly is more likely to have negative values for higher masses before the $M_{TOV}$ is reached. For the lower mass bounds, the trace anomaly after the initial fall is more or less constant and has slight variation.

 Although mass measurements are very accurate and several pulsar masses are accurately determined, the same is not valid for its radius. There have been very few observations of radius with a maximum accuracy of around 2 km. However, to constrain the EoS further, simultaneous mass and radius measurement with reasonable accuracy is needed. Also, one needs much more radius measurements of different pulsars. Presently, we are restricted to only two radius measurements of pulsars. Out of these two measurements, PSR J0030 + 0451 can constrain the EoS to a greater extent from the lower bounded region. It indicates the rejection of more softer EoS, favouring stiffer EoS. This is because of the fact that the accuracy with which it measures the radius is more precise than that of PSR J0740 + 6620.

 From the above discussion, one needs a more detailed analysis with more radius measurement for robust results. Therefore, we created an ensemble of agnostic EoS to analyze the effect of radius measurement of the EoS of NSs. In the analysis, all our EoS lies in the radius range of $10.5-14.5$ km. Therefore, first, we selected three radius windows with an accuracy of 2 km (the best possible present accuracy of NICER). As the radius window shifts towards
 the higher side, the enclosed region of the contour curves shifts towards higher pressure values, indicating a stiffening of the EoS. However, after the stiffening, the EoS also softens steeply for the higher radius window. The reversal of the nature of stiffness beyond the peak value indicates that the larger stars have softer cores, while the smaller stars have stiffer cores. It can also be seen that the equation of states producing massive stars attains their peak values at lower densities compared to those producing smaller stars. Such a trend is observed for all our mass values.

 The precision of radius measurement can vary with mass, detectors, methods and distance of the pulsar. Therefore, we also compare with varying precision in two ways: keeping the maximum radius fixed or the minimum radius fixed. The maximum precision is kept at 2 km, and the minimum is 4 km (the largest radial extent of the present analysis).

 Keeping the maximum radius fixed, we find that the EoS becomes stiffer as precision increases. However, there is again a reversal of the stiffness after the peak value is reached. Therefore, this indicates that intermediate-mass stars have very stiff cores, whereas very massive stars have relatively softer cores. However, a completely different scenario is obtained when one keeps the minimum radius fixed and increases the radial precision. This leads to a softening of the EoS, as a lower radius indicates a relatively softer EoS.

 In this work, simple parametrization of the data for radius measurement has been done. However, more complicated analysis can be done using more rigorous statistical methods of neural networks. A more rigorous method can indicate what the EoS of NS favours if future measurements of more pulsars are done. Also, improving the precision can constrain the EoS further (Appendix B); however, in the present work, we have chosen it to be a value consistent with recent NICER measurements.
 
\section*{acknowledgement}
 The authors are grateful to IISER Bhopal for providing all the research and infrastructure facilities. RM would also like to thank the SERB, Govt. of India, for monetary support in the form of Core Research Grant (CRG/2022/000663).\\

\section*{Data Availability}

This is a theoretical work; hence, it does not have any additional data. 
 
\bibliography{References}


\appendix
\section{}

Tidal deformability is an extensively studied parameter for neutron stars. It quantifies the star's deformation in response to an external tidal field. 
Employing the constraints of binary tidal deformability ($\tilde{\Lambda}$) from the event of GW170817 can help us to constrain our EoSs.

For an isolated star tidal deformability is given by $\Lambda = \frac{2}{3}k_2(\frac{R}{M})^5$, where $R$ is the radius, $M$ is the mass and $k_2$ is Love number of the stellar model \citep{Hinderer_2010}. With $\Lambda$ calculated, binary tidal deformability ($\tilde{\Lambda}$) is then given by:\par
\begin{align}
    \tilde{\Lambda} = \frac{16}{13} \frac{(12M_2 + M_1)M_1{^4}\Lambda_1 + (12M_1 + M_2)M_2{^4}\Lambda_2}{(M_1 + M_2)^5},
\end{align}
where the subscripts $1,2$ refers to the components of the binary system \citep{Leslie_Wade, Flanagan_2008, Ecker_2022}.

\begin{figure}[h]
    \includegraphics[height = 5cm, width =0.99\linewidth]{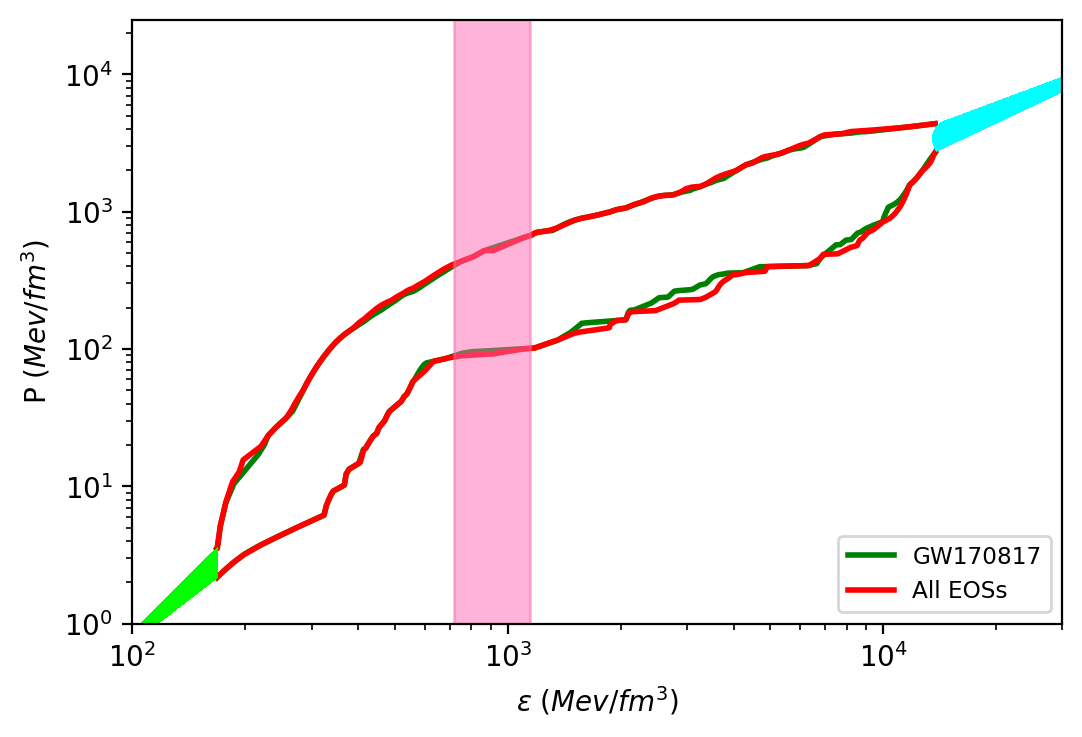}
    \caption{The contour plots of EoS comparing the original dataset (red) to the ones satisfying the GW170817 condition (green).}
    \label{6}
\end{figure}

It is now believed that $\tilde{\Lambda} < 720$, for $M_{chirp} = 1.186M_{\odot}$; where $M_{chirp}$ is the chirp mass given by $M_{chirp} ={(M_1 M_2)^{\frac{3}{5}}}{(M_1 + M_2)^{-\frac{1}{5}}}$ with mass ratio: $q = \frac{M_2}{M_1} \in [0.73, 1]$ as reported by the LIGO/Virgo data from the detection of GW170817 \cite{Abbott_2016}.

We have imposed the GW170817 condition on our dataset and shown the results in Fig \hyperref[6]{{6}}. It is evident that the constraint does not have any significant constraining effect in the $\epsilon - P$ region (once mass and radius constraints are imposed from NICER measurement); hence, we have not included its results in our main study. 

\section{}

\begin{figure}[h]
    \includegraphics[height = 5cm, width =0.99\linewidth]{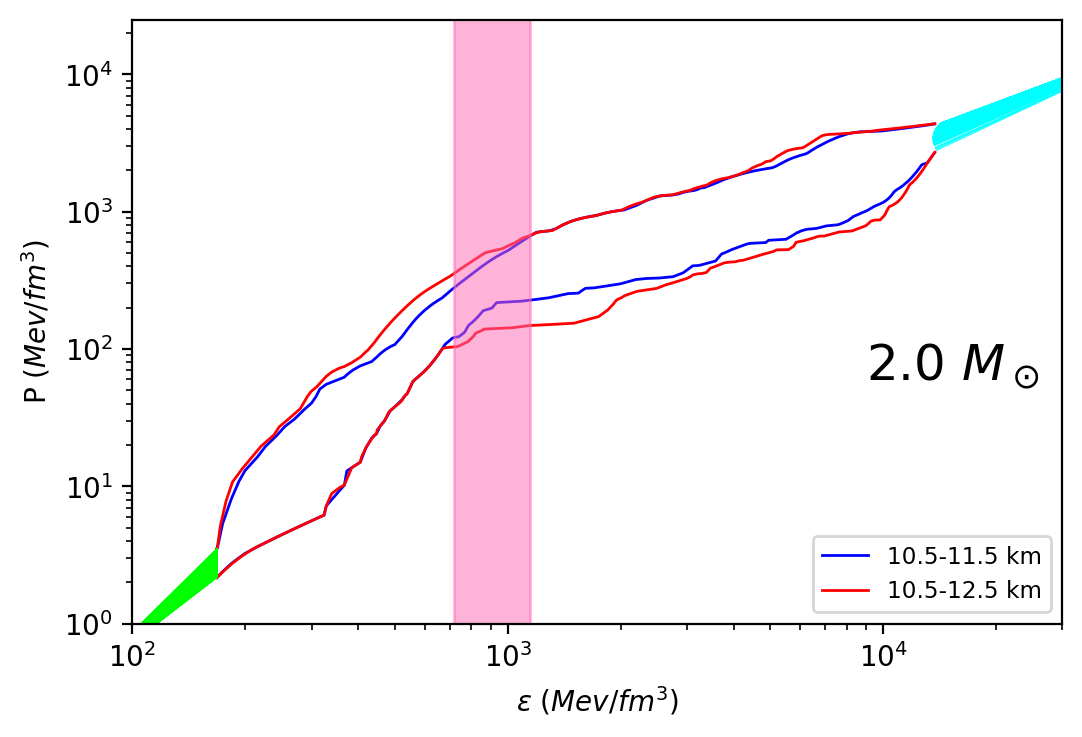}
    \includegraphics[height = 5cm, width =0.99\linewidth]{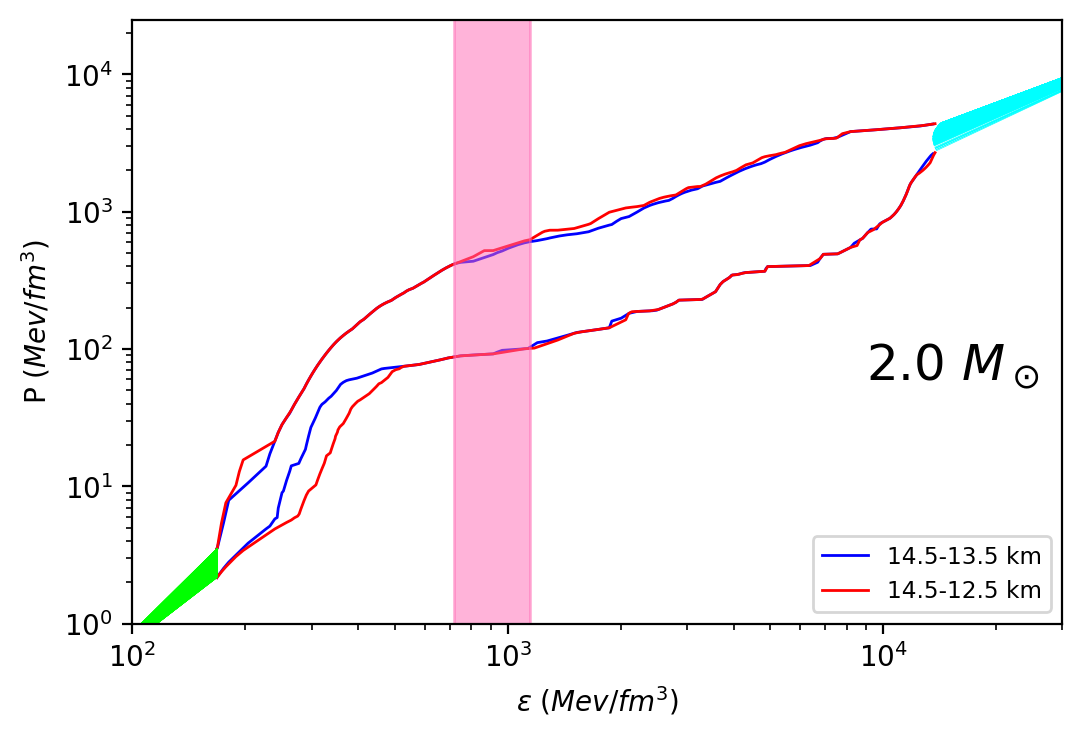}
    \label{7}
    \caption{The contour plots of EoS comparing 2km and 1km radius bounds for stellar mass 2$M_\odot$.}
\end{figure}

We have only analyzed our result with the constraint that the maximum precision for radius measurement is about 2 km. However, here we check how the bound on the EoS improves (Fig \hyperref[7]{{7}}) if we can have the precision to 1 km with future detectors. 
Keeping the lower bound fixed at 10.5 km, we find that the precision of 1 km drastically improves the result. Beyond $3-4$ times saturation density, some stiffer EoS are eliminated. Interestingly, this precision measurement also affects the density range beyond the NS density. In the density range beyond NS max density, the softer EoS are eliminated. It forms a very narrow band near the maximum density of NSs. 

In contrast, if we fix the upper bound at 14.5 km with 1 km precision, at the lower density, it eliminates most of the softer EoS; however, higher density range bounds are not much affected. Therefore, to have a clear picture of the matter properties at neutron star cores, one needs precise measurements to unravel them.

\end{document}